\documentclass[12pt]{iopart}
\usepackage{epsfig}
\usepackage{graphicx}
\usepackage{dcolumn}% Align table columns on decimal point
\usepackage{bm}% bold math

\begin{document}

% ***********************************************************************
%
\hyphenation{nuc-leon spec-tro-me-ter qua-dru-pole}

\title[Anisotropic flow at RHIC]{Anisotropic flow at RHIC:
How unique is the number-of-constituent-quark scaling?}

\author{Y. Lu${}^{1,2}$, M. Bleicher${}^3$, F. Liu${}^1$, Z. Liu${}^1$,
H. Petersen${}^3$, P. Sorensen${}^4$, H.~St\"{o}cker${}^{3,5}$, N. Xu${}^2$, X. Zhu${}^{5,6}$ \\[.4cm]}

\address{${}^1$~Institute of Particle Physics (IOPP), Central China Normal University (Huazhong Normal University), Wuhan 430079, China \\
${}^2$~Nuclear Science Division, Lawrence Berkeley National Laboratory, Berkeley, CA 94720, USA \\
${}^3$~Institut f\"ur Theoretische Physik, Johann Wolfgang Goethe-Universit\"at, Max-von-Laue-Str.~1, 60438 Frankfurt am Main, Germany \\
${}^4$~Brookhaven National Laboratory, Upton, NY 11973, USA \\
${}^5$~Frankfurt Institute for Advanced Studies (FIAS), Max-von-Laue-Str.~1, 60438 Frankfurt am Main, Germany \\
${}^6$~Physics Department, Tsinghua University, Beijing 100084, China}

\begin{abstract}
The transverse momentum dependence of the anisotropic flow $v_2$ for $\pi$, $K$, nucleon,
$\Lambda$, $\Xi$ and $\Omega$ is studied for Au+Au collisions at $\sqrt{s_{\rm NN}} =
200$~GeV within two independent string-hadron transport approaches (RQMD and UrQMD). Although
both models reach only 60\% of the absolute magnitude of the measured $v_2$, they both
predict the particle type dependence of $v_2$, as observed by the RHIC experiments: $v_2$
exhibits a hadron-mass hierarchy (HMH) in the low $p_T$ region and a
number-of-constituent-quark (NCQ) dependence in the intermediate $p_T$ region. The failure of
the hadronic models to reproduce the absolute magnitude of the observed $v_2$ indicates that
transport calculations of heavy ion collisions at RHIC must incorporate interactions among
quarks and gluons in the early, hot and dense phase. The presence of an NCQ scaling in the
string-hadron model results suggests that the particle-type dependencies observed in
heavy-ion collisions at intermediate $p_T$ might be related to the hadronic cross sections in
vacuum rather than to the hadronization process itself.
\end{abstract}

\maketitle
%--=================================================================
\section{Introduction}

The main goal of high energy nuclear collisions is the exploration of QCD-matter at high
temperatures and densities. The search for a new form of matter consisting of deconfined
quarks and gluons as constituents (usually called the Quark-Gluon-Plasma, QGP) has been a
focal point of theoretical and experimental studies over the last decade \cite{Bass:1998vz}.
Measurements of the collective motion of hadrons produced in high-energy nuclear collisions
have long been suggested as a valuable tool to gain information about the nature of the
constituents and the equation of state of the system in the early stage of the reaction
\cite{Stoecker:1981iw,Stoecker:1986ci,sorge97,ollitrault92,ritter97,Bleicher:2000sx,Zhu:2005qa,Zhu:2006fb,Mrowczynski:2002bw}.
Specifically, strange and multi-strange hadron elliptic and radial flow results seem to
indicate that the observed collectivity originates from a partonic phase\footnote{The term
partonic is used to denote any kind of deconfined matter made of quarks and gluons.}
\cite{xuinpc04}. Furthermore, elliptic and radial flow measurements for heavy-flavor hadrons
like $J/\Psi$ and D mesons will test the hypothesis of early thermalization in these
collisions \cite{Svetitsky:1996nj,Svetitsky:1997xp,xin04}.

At RHIC, measurements \cite{starklv2,phenixv2} of the elliptic flow $v_2$ and the nuclear
modification factor $R_{AA,CP}$ for identified particles have led to the conclusion that
hadrons ought to be formed via the coalescence or recombination of massive quarks
\cite{Scherer:1999qq,Bass:1998qm,Hofmann:1999jx,Hofmann:1999jy,Scherer:2001ap,Scherer:2005sr,voloshin02,rfries04,rudy04}.
A cornerstone of this conclusion is the observed number-of-constituent-quark (NCQ-) scaling
of the flow of baryons vs. mesons. Because this interpretation addresses key issues in
high-energy nuclear collisions such as deconfinement and chiral symmetry restoration, it is
of utmost importance to conduct a systematic study of other possible explanations for the
observed particle type dependencies.

For the present analysis we employ two independent hadron-string transport models RQMD(v2.4)
and UrQMD(v2.2) \cite{rqmd,urqmd22} to study the effect of hadronic cross sections,
kinematics etc. on the particle type dependence of $v_2$. We present the $v_2$ values of
$\pi$, $K$, $p$, $\Lambda$, $\Xi$ and $\Omega$ for Au+Au collisions at $\sqrt{s_{\rm NN}} =
200$~GeV. We find that - although the hadronic models give only 60\% of the magnitude of
$v_2$ - both models reproduce the gross features of the particle type dependence, including
the mass ordering at $p_T < 1$~GeV/c and the number-of-constituent-quark dependence at $p_T >
1$~GeV/c. Within both models, the NCQ dependence is related to the additive-quark-model for
hadronic cross-sections. These findings imply that detailed comparisons between data and
theory are necessary to disentangle hadronic and partonic effects at intermediate transverse
momenta, $1.5$~GeV/c~$< p_T <~5$~GeV/c. For a transport model discussion of the quantitative
accuracy and validity of the various methods to extract elliptic flow values from the data,
the reader is referred to \cite{Bleicher:2000sx,Zhu:2005qa,Zhu:2006fb}

\begin{figure}[t]
\begin{center}
\begin{minipage}[b]{6cm}\includegraphics[width=6cm]{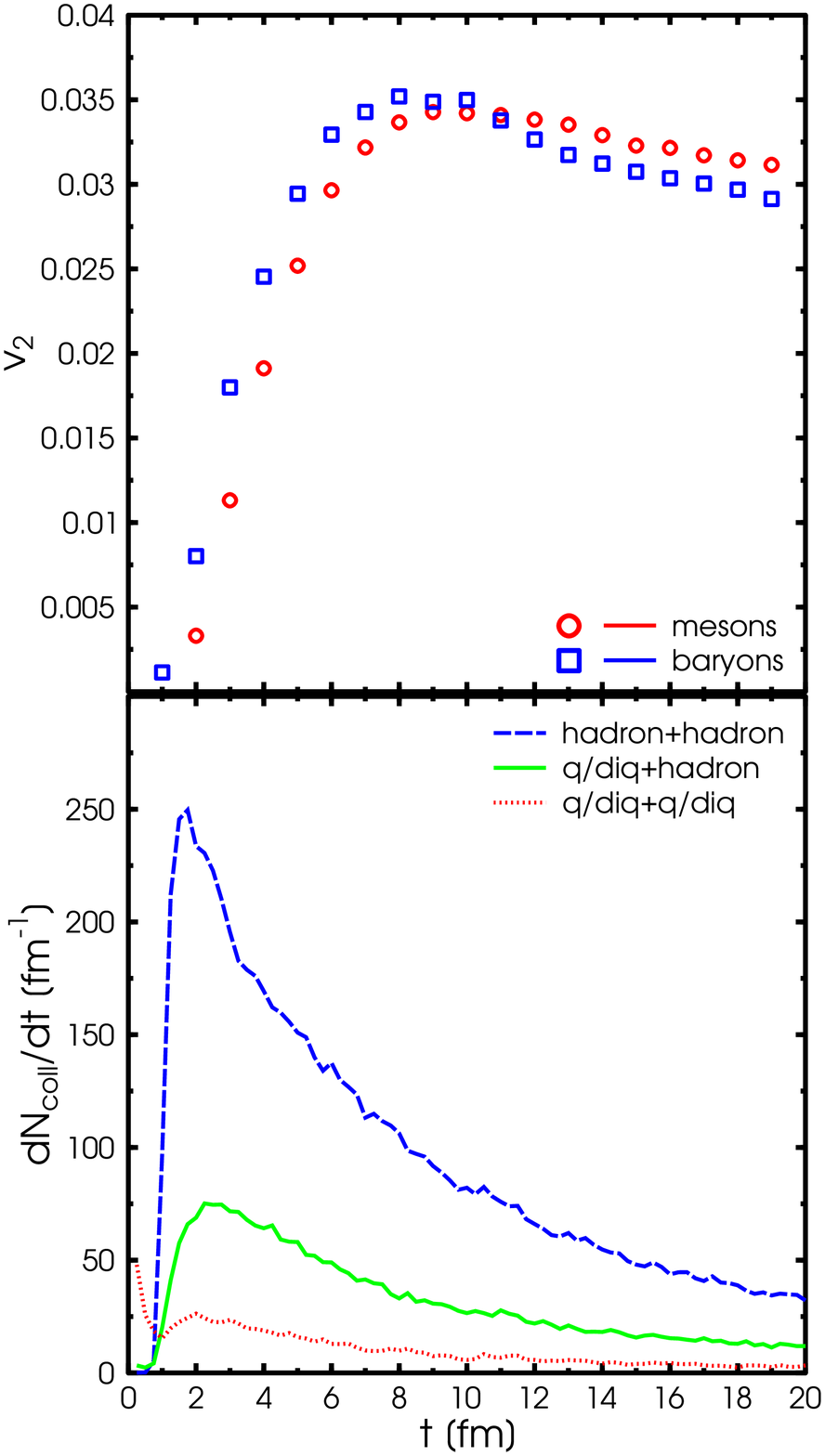}
\end{minipage}
\begin{minipage}[b]{6cm}
\includegraphics[width=6cm]{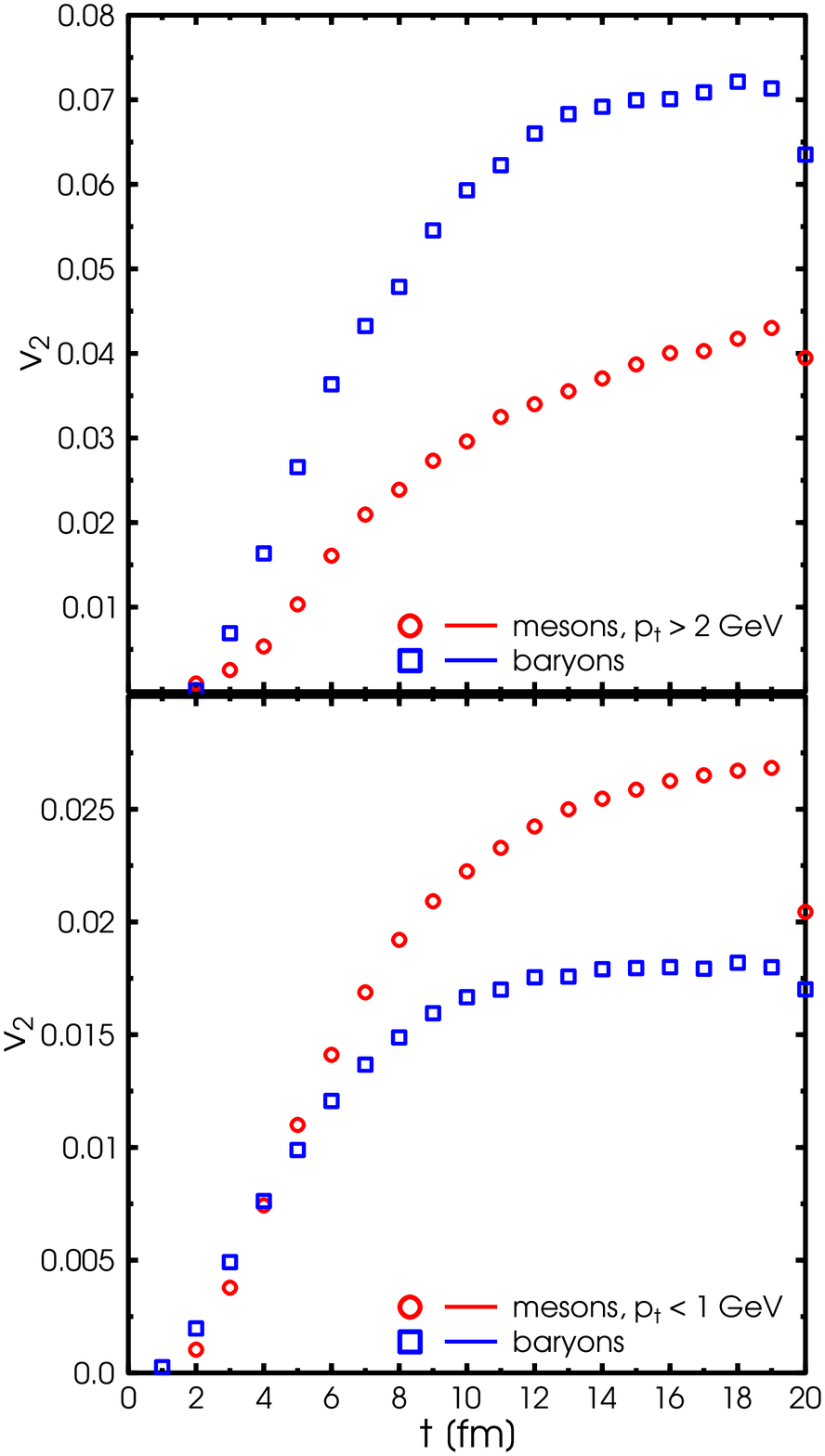}
\end{minipage}
\end{center}
\caption[]{\label{fig0} UrQMD calculation for Au+Au interactions at $\sqrt{s_{\rm NN}} =
200$~GeV. Top left: Meson and baryon $v_2$ versus time at midrapidity. Bottom left: Collision
rates as a function of time for hadron+hadron (dashed line), hadron+quark/di-quark (full
line) and quark/di-quark+quark/di-quark (dotted line). Bottom right: Meson and baryon $v_2$
versus time at midrapidity and  $p_T\le 1$~GeV/c. Top right: Meson and baryon $v_2$ versus
time at midrapidity and  $p_T\ge 2$~GeV/c.}
\end{figure}

As one can observe from Fig. \ref{fig0} the build-up of the elliptic flow in these models is
tightly connected to the hadron+hadron scattering rate (top/bottom figures on the lhs.).
Initial valence (di-)quark scatterings before 1~fm/c do only yield a negligible amount of
$v_2$. The right panels in Fig. \ref{fig0} show the time evolution of the elliptic flow for
mesons and baryons for low $p_T$ (bottom) and high $p_T$ (top). Already at early times, a
separation between meson and baryon elliptic flow is visible, leading to the mass hierarchy
at low transverse momenta and to the stronger elliptic flow of baryons (compared to mesons)
at high $p_T$.

%%%%%%%%%%%%%%%%%%%%%%%%%%%%%%%%

\section{Model Results}

Within the framework of the hadronic transport approach, a typical heavy ion collision
proceeds schematically in three stages, i.e. the pre-hadronic (strings and constituent
(di-)quarks) stage, the hadronic pre-equilibrium stage, the evolution towards hadronic
kinetic equilibrium and freeze-out. The pre-hadronic stage  involves the initial excitation
and fragmentation of color strings and ropes. At the highest RHIC energy, this stage lasts
for about 0.5-1.5~fm/c and the effective transverse pressure/EOS is rather soft. During the
late hadronic stage, the hadronic system approaches local kinetic equilibrium followed by an
approach to free-streaming, where the system escapes equilibrium due to dilution of the
hadronic gas: the mean free path of the hadrons exceeds the finite size of the system
\cite{rqmd,urqmd22,rqmd2}, the free streaming hadrons decay and feed down to the lightest
species.

\begin{figure}[t]
\centerline{\includegraphics[height=8cm]{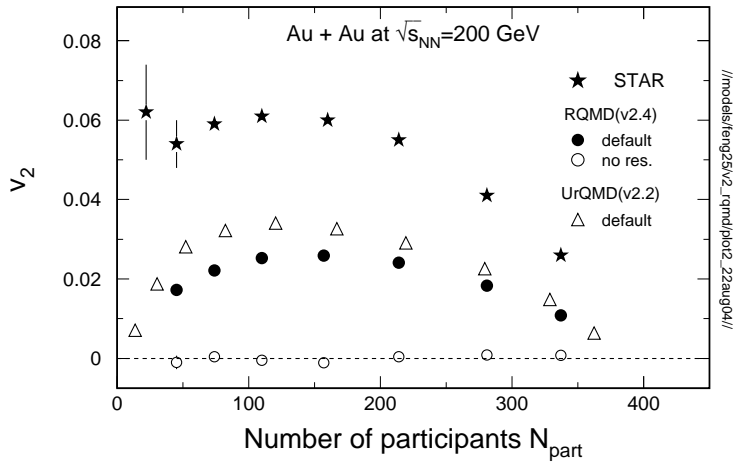}}
\vspace{-0.25cm}
\caption[]{\label{fig1}Charged hadron $v_2$ versus the number of
  participating nucleons in Au+Au interactions at $\sqrt{s_{\rm NN}} = 200$~GeV.
 Experimental data \cite{star1v2,starv204} from the 4-particle cumulant method
  is shown as stars. RQMD results are depicted by filled circles (full calculation) and
  open circles (without rescattering). The UrQMD calculations are shown as open triangles.}
\end{figure}

Figure \ref{fig1} presents the model results on the centrality dependence of the charged
hadron $v_2$-values along with measurements from the STAR collaboration
\cite{star1v2,starv204}. Both hadronic transport models (UrQMD v2.2 and RQMD v2.4) reach
about 60\% of the measured $v_2$ values only, although the centrality dependencies are very
similar to the data. There is a small variance between the two models, which we consider as
an estimate of the systematic errors in such model calculations. Although the $v_2$ values
from the hadronic transport model also depend on the formation-time of hadrons from strings
\cite{Bleicher:2000sx}, the failure of both hadronic transport models to describe
quantitatively the magnitude of $v_2$ is a strong indication that there are interactions
amongst pre-hadronic constituents (partons) present in nature (but not in the hadron-string
models discussed here), which are responsible for the large $v_2$ values observed in the
experiments \cite{Burau:2004ev,Xu:2005tm,Bratkovskaya:2004kv}. When rescattering between the
hadrons is turned off (full circles), $v_2$ vanishes completely, because repulsive vector
interactions are not included into the present simulations
\cite{Stoecker:1981iw,Stoecker:1986ci,Bleicher:2000sx}.

\begin{figure}[t]
\centerline{\includegraphics[height=8cm]{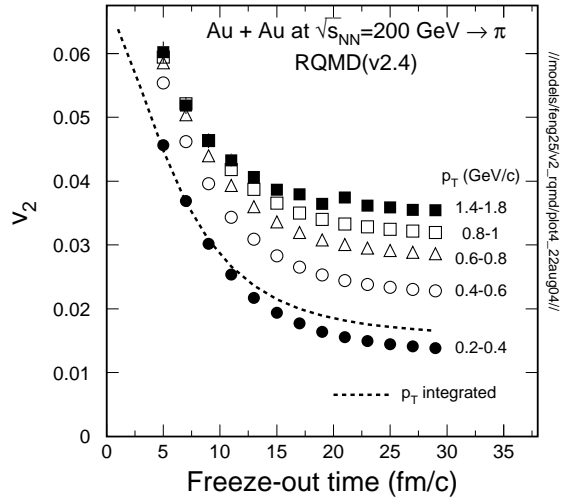}}
\vspace{-.25cm}
\caption[]{\label{fig2}Elliptic flow ($v_2$) of pions from RQMD versus their freeze-out time for several $p_T$
  windows for  minimum bias Au+Au collisions at $\sqrt{s_{\rm NN}}= 200$~GeV.
 The dashed-line represents the $p_T$ integrated
  distribution.}
\end{figure}

Let us now analyze the temporal structure of the elliptic flow's development.
Figure \ref{fig2} shows the $v_2$ values as a function of the  freeze-out time for
pions in minimum biased Au+Au interactions at $\sqrt{s_{\rm NN}} = 200$~GeV.
The different symbols denote different transverse momentum intervals
decreasing from top to bottom.
One clearly observes a strong correlation between freeze-out time and elliptic flow:
particles that decouple earlier have a larger $v_2$ value than
those that freeze-out later. The lower $p_T$ particles tend to
freeze-out later and their $v_2$ continues to evolve late in the
evolution (a similar correlation has been observed earlier in low energy collisions \cite{Bass:1994af})
The decrease in the $v_2$ towards later times
is related to the reduction of the coordinate-space anisotropy with time - i.e. the system
has become more spherical than it was at earlier times. The $v_2$ values of higher
$p_T$-pions saturate sooner and tend to reflect the earlier stage of
the collision more strongly. Thus, within the model dynamics, the final $v_2$ is mostly driven by the
early stages of the reaction; the $v_2$ values at high $p_T$ are closer to the initial/early $v_2$ values than the $v_2$ values at lower $p_T$.

\section{Particle Type Dependence}

How much of the observed NCQ-scaling features can be reproduced by the hadronic models? In both dynamical approaches, finite (vacuum) cross sections are used to model the strong interactions in the
hadron-string cascade. Unlike the simplistic Cooper-Frye freeze-out treatment in most hydrodynamic
calculations, the transition from strongly interacting matter to free-streaming is determined here by the interplay of the local particle
density and the energy dependent cross section of the individual hadrons.
It is well known that a proper treatment of the gradual freeze-out is crucial for the
finally observed hadron distributions. It was pointed out that the hydrodynamical
results on flow depend strongly on the proper kinetic treatment of the freeze-out process and can not be
approximated by isotherms \cite{Hung:1997du,Anderlik:1998et,Magas:1999yb}.

\begin{figure}[t]
\centerline{\includegraphics[height=10cm]{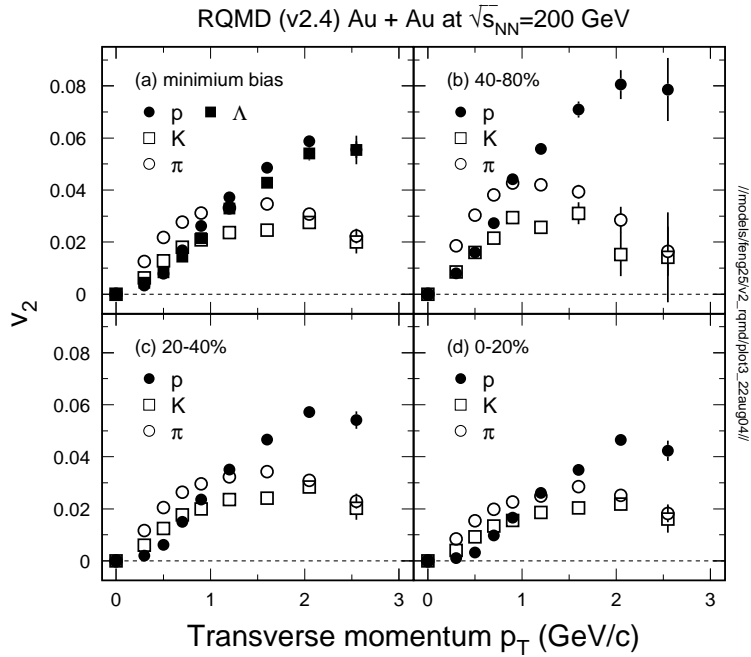}}
\vspace{-.25cm}
\caption[]{RQMD results of $\pi$, $K$, $p$, and $\Lambda$ $v_2$ from Au+Au collisions at $\sqrt{s_{\rm NN}} =
200$~GeV. (a) Minimum bias collisions: At about $p_T$ $\sim$ 1.2 GeV/c, baryon and meson $v_2$ are crossing
each other; (b) 40-80\%; (c) 20-40\%; (d) 0-20\%.}

\label{fig3}
\end{figure}

However, the major shortcoming of the present hadron-string approach is the lack of the early
partonic interactions which are important for the early dynamics in ultra-relativistic heavy
ion collisions \cite{Bass:1998vz,Stoecker:1981iw,Stoecker:1986ci,Bleicher:2000sx,miklos05}.
In order to take care of both partonic and hadronic interactions in high-energy nuclear
collisions, a combination of the hydrodynamic model for the early stage dynamics (the
``perfect'' fluid stage) with a hadronic transport model for the later stage plus freeze-out
has been proposed
\cite{Bass:1999tu,Dumitru:1999sf,Bass:2000ib,teaney01,Nonaka:2005aj,hirano05}. Fig.
\ref{fig3} shows the collision centrality dependence of the $p_T$-dependent $v_2$ values for
$\pi$, $K$, $p$, and $\Lambda$. Both, the hydrodynamic behavior (in the low $p_T$ region) and
a hadron-type dependence (in the intermediate $p_T$ region) are clearly predicted in all
centrality bins. This ``crossing and subsequent splitting" between meson- and baryon elliptic
flow as well as the breakdown of the hydrodynamical mass scaling at high transverse momenta
was first predicted within the UrQMD model \cite{Bleicher:2000sx} and has later been observed
in the experimental data. It is important to note that the more recent explanations of this
effect (the suggested ``number-of-constituent-quark" scaling) is not a unique feature of the
``quark recombination/coalescence" assumption: hadronic interactions alone have
quantitatively (at the correct $p_T$-values) predicted this hadron type dependence.

\begin{figure}[t]
\centerline{\includegraphics[height=8cm]{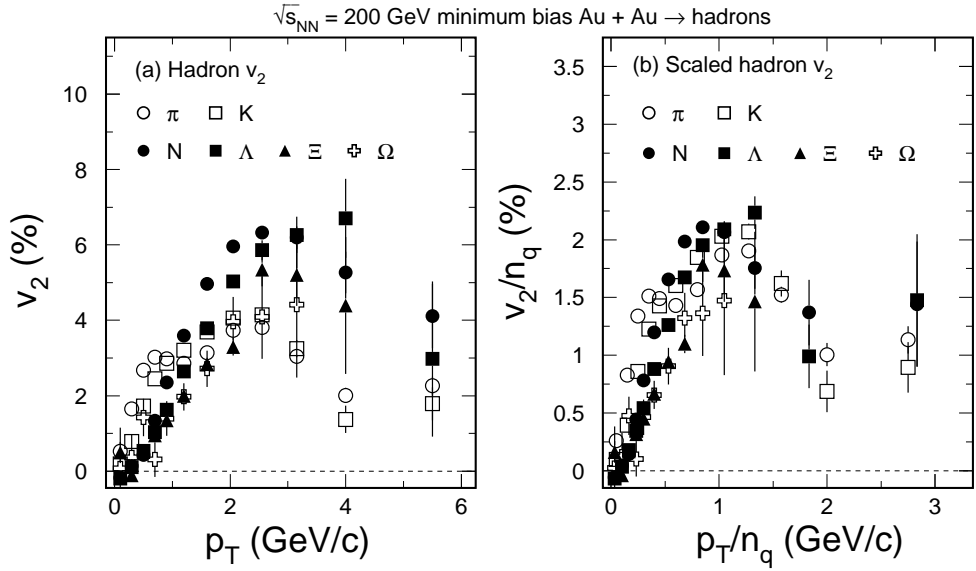}} \vspace{-0.25cm}
    \caption[]{(a) Hadron $v_2$ from minimum bias Au+Au collisions at
    $\sqrt{s_{\rm NN}}= 200$~GeV; (b) Scaled hadron $v_2$ results are shown. The
    $n_q$ refers to the number of constituent quarks. Symbols
    represent results from the UrQMD (v2.2) model for various hadron species.
    At low $p_T/N_q \le 0.5$ GeV/c, $\pi$
    does not follow the scaling perhaps caused
    by the resonance decay \cite{xin04}.
    In higher $p_T$ region, $K$ meson seems to fall off the scaling curve due to the comparatively small hadronic cross sections in the model.}
\label{fig4}
\end{figure}

Let us explore the $p_T$-dependence of the event anisotropy parameters in  detail. Figure
\ref{fig4} shows the calculated unscaled and scaled $v_2$ values of various hadrons versus
the unscaled and scaled transverse momenta, $p_T$, of the various hadrons. On the left hand
side (Figure \ref{fig4}(a)) one can see that at lower transverse momenta, $p_T$ $\le$ 1.5
GeV/c, the heavier hadrons exhibit smaller $v_2$ values than the lighter hadrons: Hadron
transport theory predicts mass ordering. The $\Xi$ and $\Omega$ $v_2$ values from the UrQMD
calculations are also included. They are the lowest of all $v_2$ values for $p_T \le 2$
GeV/c. Such mass ordering is exactly what is observed in the experimental data
\cite{starklv2} and is in accord with hydrodynamic calculations \cite{pasi01}. Hence,
hadronic interactions, which do take place at later stages of the collisions, also do
contribute to the observed collective motion.

At higher $p_T$ values, this mass dependence gives way to the $v_2(p_T)$-dependence on the
hadron type (i.e. meson or baryon). Here, it is interesting to note that the $\Omega$-baryons
seemingly acquire a significant amount of $v_2$ in the model calculations. In addition, there
is also clear, but small difference for kaons and pions in $v_2$ values at $p_T \ge
1.5$~GeV/c. This particle type dependence, rather than the otherwise dominating particle mass
dependence, is also observed in the data \cite{starklv2}. It is important to note that the
$\phi$ meson has a mass that is very close to the mass of the baryons $p$ and $\Lambda$, and,
indeed, recent experimental results on the $\phi$'s $v_2$ values are similar to other mesons
\cite{markusqm05}. However, in the hadronic transport model, about 2/3 of the $\phi$-mesons
are formed via $K$-$\bar{K}$-coalescence, which is not necessarily the dominant process in
heavy ion collisions \cite{star1234}. Therefore, the $v_2$ values of $\phi$ meson are not
shown in Figure \ref{fig4}. It should also be noted that in the high $p_T$ region, $p_T \ge
2.5$~GeV/c, all $v_2$ values start to decrease. This indicates that the system is deviating
from an ideal hydrodynamic behavior.  This trend is best seen in the right, ``scaled", plot
in Fig. \ref{fig4}. Such a drop has been observed in the data.
%however, there it starts at about $p_T \approx 5$~GeV/c. The difference in this onset-$p_T$ may be attributed to the apparent partonic collectivity in heavy ion collisions at RHIC.

The test of the NCQ scaling hypothesis is shown in Figure \ref{fig4}(b), which depicts the
scaled hadron values, $v_2/n_q$. The scaling factor is the number of constituent quarks (NCQ)
in accord with the coalescence approach \cite{voloshin02,rfries04,rudy04}. For mesons and
baryons, $n_q=2$ and $n_q$=3, respectively. The NCQ-scaling is clearly observed in both RQMD
(not shown here) and UrQMD model calculations except for the pions. This surprising result
and its implications for the frequently invoked recombination/coalescence hypothesis will be
discussed in the last section.

%%%
However, one should note that there are experimentally distinguishable differences (when
differential $\Omega$ elliptic flow data with good statistics becomes available) between
``real'' NCQ scaling from coalescence and the approximate NCQ scaling due to the cross
section hierarchy discussed here. In the coalescence approach one expects  identical elliptic
flow of all baryons irrespective of their type. In contrast,  the present AQM scaling
picture, leads to an observable
 ordering of the elliptic flow at fixed $p_T$ according to the strangeness content. This can
indeed be observed in Fig. \ref{fig4}, where an ordering of the elliptic flow values can be observed:
$v_2({\rm N})>v_2(\Lambda)>v_2(\Xi)>v_2(\Omega)$.
%%%%%

\section{Discussion and Conclusions}
%The quark coalescence or quark recombination mechanism for hadronization is a rather general
%idea, even applicable to elementary collisions. For example, when one considers the string
%picture in $e^+e^-$ collisions, hadrons are formed by {\it coalescence} of quarks from a
%fragmenting color flux tube. In high energy nuclear collisions, the local parton density can
%be much higher than that in elementary collisions. There, hadrons can also be formed, in
%principle, via {\it coalescence} of quarks from different strings \cite{Scherer:2001ap}. The
%process of string fusion (or color rope formation) is well known in nuclear collisions and
%has been studied in detail in \cite{Biro:1984,Mohring:1992wm,Sorge:1995dp}. These
%multi-parton processes can therefore be manifested in the observed hadron momentum
%distributions \cite{erice03} even in transport models without an explicit assumption of QGP
%formation.

The particle and energy density is highest at the center of the created fireball in
relativistic nuclear collisions - initially, there is an angular dependent matter density
gradient. The repulsive interactions among the constituents will therefore push matter to
move outwards. In this way, the collective flow develops in nucleus-nucleus collisions
\cite{Bass:1998vz,Stoecker:1981iw,Stoecker:1986ci,sorge97,ollitrault92,ritter97}. We would
like to stress that flow means matter- and energy flow. It is independent of the type of
particles, either partons or hadrons, or different kinds of hadrons. Hence, by studying the
collective motion of the produced hadrons one can, in principle, extract the information of
early collision dynamics \cite{sorge97,ollitrault92,erice03}. In general, one expects that
the final elliptic flow,
\begin{equation}
v_2(p_T) \propto \int_t \int_{\Sigma} \sigma(\rho,p_T) \otimes
\rho_{\Sigma}(t,x,y,p_T) d\vec{A}_{\Sigma}(x,y)dt,
\label{eqv2}
\end{equation}
where $\Sigma$ denotes the hyper-surface where hadrons are emitted, will depend on $\sigma$,
i.e. the interaction cross section, which, in principle, depends on the particle type, cm
angle and relative momenta. The specific particle density depends on the collision time $t$,
location, and momentum. For short mean free paths, the transverse flow is intimately related
to the pressure, which in turn depends on the density and temperature of the matter under
study
\cite{Bass:1998vz,Stoecker:1981iw,Stoecker:1986ci,sorge97,ollitrault92,ritter97,Bleicher:2000sx}.
Indeed, the frequent rescatterings among the hadrons can lead to hydrodynamic-like mass
ordering in the low $p_T$ region.

At the higher transverse momenta, $p_T\ge 1.5$~GeV/c, the particles escape quickly from the
system to low density, in effect leading to early freeze-out and lack of development of the
hydrodynamics and the details of the interaction cross-sections are most important. As the
cross sections depend on the particle type, for mesons or baryons to first approximation
given by the constituent quark model \cite{voloshin02}, we do expect roughly a 2:3 scaling of
the meson-to-baryon elliptic flow from transport calculations.
%In this sense, the observation
%of NCQ scaling in $v_2$ may represent a rediscovery and confirmation of the additive quark
%model for hadron cross sections.

The hadronic models underpredicted the strength of $v_2$ at RHIC, because
early partonic interactions (except from quark coalescence during the string break-up)
are not included in the model. The early stage with highest density and smallest mean free paths is
``missing". This shortcoming of the
hadronic models clearly demonstrates the need for the early, dense partonic
interactions in heavy ion collisions at RHIC.

%--=================================================================
\section{Summary}

In summary, the hadronic transport models UrQMD and RQMD have been employed to study the
elliptic flow of hadrons in Au+Au collisions at the highest RHIC energy. We have analyzed the
$v_2$ values as a function of collision centrality, transverse momenta and collision time for
various meson- and baryon species, including multi-strange baryons. Both hadron-string
transport models fail by 40\% to exhaust the absolute value of $v_2$, probably due to the
lack of (partonic) interactions in the initial, hot and dense stage. However, because of the
hadronic re-interactions, the hadron mass hierarchy is qualitatively well reproduced in the
low $p_T$ region. Rescattering is the key that leads to the quasi-ideal hydrodynamic
appearance in $v_2(p_T)$. Also in the intermediate $p_T$ region, the hadron type dependence
(number-of-constituent-quark scaling) is predicted by both hadronic transport models. Here,
this dependence is due to the hadronic cross sections which do roughly scale with the number
of constituent quarks, in accord with the additive quark model. This finding challenges the
recent interpretation of NCQ scaling as a unique deconfinement signature. Thus, further tests
of the deconfinement-plus-recombination hypothesis are necessary with high precision $v_2$
measurements of resonance hadrons like $K^*$, $\rho$, $\Delta$, $\Lambda^*$, and $\Xi^*$.

%--===================================================================
\section*{Acknowledgements}

The authors would like to thank C.~M. Ko, A. Poskanzer, H.~G. Ritter, and K. Schweda for
enlightening discussions. This work was
supported by BMBF and GSI, Germany, by the HENP Divisions of the Office of
Science of the U.S. DOE and by the NSFC of China under the project 10575042.\\[.5cm]

%--=================================================================

%--=================================================================

\section{References}
\begin{thebibliography}{999}

%\cite{Bass:1998vz}
\bibitem{Bass:1998vz}
  S.~A.~Bass, M.~Gyulassy, H.~St\"ocker and W.~Greiner,
  %``Signatures of quark-gluon-plasma formation in high energy heavy-ion
  %collisions: A critical review,''
  J.\ Phys.\ G {\bf 25}, R1 (1999)
  [arXiv:hep-ph/9810281].
  %%CITATION = HEP-PH 9810281;%%

%\cite{Stoecker:1981iw,Stoecker:1986ci,sorge97,ollitrault92,ritter97,Mrowczynski:2002bw}
\bibitem{Stoecker:1981iw}
  H.~St\"ocker, M.~Gyulassy and J.~Boguta,
  %``Probing Dense Nuclear Matter Via Nuclear Collisions,''
  Phys.\ Lett.\ B {\bf 103}, 269 (1981).
  %%CITATION = PHLTA,B103,269;%%

%\cite{Stoecker:1981iw,Stoecker:1986ci,Bleicher:2000sx}
\bibitem{Stoecker:1986ci}
  H.~St\"ocker and W.~Greiner,
  %``High-Energy Heavy Ion Collisions: Probing The Equation Of State Of Highly
  %Excited Hadronic Matter,''
  Phys.\ Rept.\  {\bf 137}, 277 (1986).
  %%CITATION = PRPLC,137,277;%%

\bibitem{sorge97} H. Sorge, Phys. Rev. Lett. {\bf 78}, 2309(1997).

\bibitem{ollitrault92} J.-Y. Ollitrault, Phys. Rev. {\bf D46}, 229(1992).

\bibitem{ritter97} W. Reisdorf and H.G. Ritter,
Ann. Rev. Nucl. Part. Sci. {\bf 47}, 663(1997).

%\cite{Bleicher:2000sx}
\bibitem{Bleicher:2000sx}
  M.~Bleicher and H.~St\"ocker,
  %``Anisotropic flow in ultra-relativistic heavy ion collisions,''
  Phys.\ Lett.\ B {\bf 526}, 309 (2002)
  [arXiv:hep-ph/0006147],
  %%CITATION = HEP-PH 0006147;%%
  H.~St\"ocker,
  %``Collective Flow signals the Quark Gluon Plasma,''
  Nucl.\ Phys.\ A {\bf 750}, 121 (2005)
  [arXiv:nucl-th/0406018].
  %%CITATION = NUCL-TH 0406018;%%
  %%Cited 41 time in SPIRES-HEP
%\cite{Zhu:2005qa,Zhu:2006fb}
\bibitem{Zhu:2005qa}
  X.~Zhu, M.~Bleicher and H.~St\"ocker,
  %``Elliptic flow analysis at RHIC: Fluctuations vs. non-flow effects,''
  Phys.\ Rev.\ C {\bf 72}, 064911 (2005)
  [arXiv:nucl-th/0509081].
  %%CITATION = NUCL-TH 0509081;%%

%\cite{Zhu:2006fb}
\bibitem{Zhu:2006fb}
  X.~Zhu, M.~Bleicher and H.~St\"ocker,
  %``Elliptic Flow Analysis at RHIC with the Lee-Yang Zeroes Method in a
  %Relativistic Transport Approach,''
  arXiv:nucl-th/0601049.
  %%CITATION = NUCL-TH 0601049;%%

%\cite{Mrowczynski:2002bw}
\bibitem{Mrowczynski:2002bw}
  S.~Mrowczynski and E.~V.~Shuryak,
  %``Elliptic flow fluctuations,''
  Acta Phys.\ Polon.\ B {\bf 34}, 4241 (2003)
  [arXiv:nucl-th/0208052].
  %%CITATION = NUCL-TH 0208052;%%

\bibitem{xuinpc04} N. Xu, Nucl. Phys. {\bf A751}, 109(2005).

%\cite{Svetitsky:1996nj,Svetitsky:1997xp}
\bibitem{Svetitsky:1996nj}
  B.~Svetitsky and A.~Uziel,
  %``Passage of charmed particles through the mixed phase in high-energy
  %heavy-ion collisions,''
  Phys.\ Rev.\ D {\bf 55}, 2616 (1997)
  [arXiv:hep-ph/9606284].
  %%CITATION = HEP-PH 9606284;%%

%\cite{Svetitsky:1997xp}
\bibitem{Svetitsky:1997xp}
  B.~Svetitsky and A.~Uziel,
  %``Charm: A thermometer of the mixed phase,''
  arXiv:hep-ph/9709228.
  %%CITATION = HEP-PH 9709228;%%

\bibitem{xin04} X. Dong, {\it et al.}, Phys. Lett. {\bf B597}, 328(2004).

\bibitem{starklv2}  J. Adams {\it et al.}, (STAR Collaboration), Phys. Rev.
Lett. {\bf 92}, 052302(2004).

\bibitem{phenixv2} S.S. Adler {\it et al.}, (PHENIX Collaboration),
Phys. Rev. Lett. {\bf 91}, 182301(2003).

%\cite{Scherer:1999qq}
\bibitem{Scherer:1999qq}
  S.~Scherer {\it et al.},
  %``Critical review of quark gluon plasma signatures,''
  Prog.\ Part.\ Nucl.\ Phys.\  {\bf 42}, 279 (1999).
  %%CITATION = PPNPD,42,279;%%
  %%Cited 25 times in SPIRES-HEP

%\cite{Bass:1998qm}
\bibitem{Bass:1998qm}
  S.~A.~Bass {\it et al.},
  %``Reaction dynamics in Pb + Pb at the CERN/SPS: From partonic degrees  of
  %freedom to freeze-out,''
  Prog.\ Part.\ Nucl.\ Phys.\  {\bf 42}, 313 (1999)
  [arXiv:nucl-th/9810077].
  %%CITATION = NUCL-TH 9810077;%%
  %%Cited 4 times in SPIRES-HEP

%\cite{Hofmann:1999jx}
\bibitem{Hofmann:1999jx}
  M.~Hofmann, M.~Bleicher, S.~Scherer, L.~Neise, H.~St\"ocker and W.~Greiner,
  %``Statistical mechanics of semi-classical colored objects,''
  Phys.\ Lett.\ B {\bf 478}, 161 (2000)
  [arXiv:nucl-th/9908030].
  %%CITATION = NUCL-TH 9908030;%%
  %%Cited 11 time in SPIRES-HEP

%\cite{Hofmann:1999jy}
\bibitem{Hofmann:1999jy}
  M.~Hofmann, J.~M.~Eisenberg, S.~Scherer, M.~Bleicher, L.~Neise, H.~St\"ocker and W.~Greiner,
  %``Nonequilibrium dynamics of a hadronizing quark-gluon plasma,''
  arXiv:nucl-th/9908031.
  %%CITATION = NUCL-TH 9908031;%%
  %%Cited 6 times in SPIRES-HEP

%\cite{Scherer:2001ap}
\bibitem{Scherer:2001ap}
  S.~Scherer, M.~Hofmann, M.~Bleicher, L.~Neise, H.~St\"ocker and W.~Greiner,
  %``Microscopic coloured quark-dynamics in the soft non-perturbative  regime:
  %Description of hadron formation in relativistic S + Au  collisions at CERN,''
  New J.\ Phys.\  {\bf 3}, 8 (2001)
  [arXiv:nucl-th/0106036].
  %%CITATION = NUCL-TH 0106036;%%
  %%Cited 5 times in SPIRES-HEP

%\cite{Scherer:2005sr}
\bibitem{Scherer:2005sr}
  S.~Scherer and H.~St\"ocker,
  %``Multifragmentation, clustering, and coalescence in nuclear collisions,''
  arXiv:nucl-th/0502069.
  %%CITATION = NUCL-TH 0502069;%%
  %%Cited 0 times in SPIRES-HEP

\bibitem{voloshin02} D. Molnar and S. Voloshin, Phys. Rev. Lett. {\bf
91}, 092301(2003).

%recombination review

\bibitem{rfries04} R. J. Fries, J. Phys. {\bf G31}, S379(2005);
nucl-th/0410085 and references therein.

\bibitem{rudy04} R. Hwa and C.B. Yang, Phys. Rev. {\bf C70},
024904(2004) and reference therein.

\bibitem{rqmd} RQMD(v2.4): H. Sorge, Phys. Rev. {\bf C52}, 3291(1995).

\bibitem{urqmd22} UrQMD(v2.2): %\cite{Bleicher:1999xi}
%\bibitem{Bleicher:1999xi}
  M.~Bleicher {\it et al.},
  %``Relativistic hadron hadron collisions in the ultra-relativistic quantum
  %molecular dynamics model,''
  J.\ Phys.\ G {\bf 25}, 1859 (1999)
  [arXiv:hep-ph/9909407],
  %%CITATION = HEP-PH 9909407;%%
S.A. Bass, {\it et al.}, Prog. Part. Nucl. Phys. {\bf
41}, 225(1998); M. Bleicher, {\it et al.}, in preparation, (2006).

\bibitem{rqmd2} H.van Hecke, H.Sorge, and N. Xu, Phys. Rev. Lett.
  {\bf 81}, 5764(1998).

\bibitem{star1v2} K.H. Ackermann, {\it et al.} (STAR Collaboration),
Phys. Rev. Lett. {\bf 86}, 402(2001).  %nucl-ex/0009011.

\bibitem{starv204} J. Adams, {\it et al.} (STAR Collaboration),
Phys. Rev. {\bf C72}, 014904(2005).  %nucl-ex/0409033.

%\cite{Burau:2004ev,Xu:2005tm}
\bibitem{Burau:2004ev}
G.~Burau, J.~Bleibel, C.~Fuchs, A.~Faessler, L.~V.~Bravina and E.~E.~Zabrodin,
%``Anisotropic flow of charged and identified hadrons in the quark-gluon
%string model for Au + Au collisions at s(NN)**(1/2) = 200-GeV,''
Phys.\ Rev.\ C {\bf 71}, 054905 (2005) [arXiv:nucl-th/0411117].
%%CITATION = NUCL-TH 0411117;%%
%%Cited 3 times in SPIRES-HEP

%\cite{Xu:2005tm}
\bibitem{Xu:2005tm}
Z.~Xu and C.~Greiner,
%``Thermalization of gluons at RHIC including g g <--> g g g interactions
in a
%parton cascade,''
Proceedings of the Quark Matter 2005, Aug. 4 - 9, 2005, Budapest, arXiv:hep-ph/0509324.
%%CITATION = HEP-PH 0509324;%%
%%Cited 1 time in SPIRES-HEP


%\cite{Bratkovskaya:2004kv}
\bibitem{Bratkovskaya:2004kv}
  E.~L.~Bratkovskaya {\it et al.},
  %``Strangeness dynamics and transverse pressure in relativistic nucleus
  %nucleus collisions,''
  Phys.\ Rev.\ C {\bf 69}, 054907 (2004)
  [arXiv:nucl-th/0402026].
  %%CITATION = NUCL-TH 0402026;%%

%\cite{Bass:1994af}
\bibitem{Bass:1994af}
  S.~A.~Bass, C.~Hartnack, H.~Stoecker and W.~Greiner,
  %``High p(T) pions as probes of the early dense reaction phase in heavy ion
  %collisions at 1-GeV/nucleon,''
  Phys.\ Rev.\ C {\bf 50}, 2167 (1994).
  %%CITATION = PHRVA,C50,2167;%%

%\cite{Hung:1997du}
\bibitem{Hung:1997du}
  C.~M.~Hung and E.~V.~Shuryak,
  %``Equation of state, radial flow and freeze-out in high energy heavy ion
  %collisions,''
  Phys.\ Rev.\ C {\bf 57}, 1891 (1998)
  [arXiv:hep-ph/9709264].
  %%CITATION = HEP-PH 9709264;%%

%\cite{Anderlik:1998et,Magas:1999yb}
\bibitem{Anderlik:1998et}
C.~Anderlik {\it et al.},
%``Freeze out in hydrodynamical models,''
Phys.\ Rev.\ C {\bf 59}, 3309 (1999)
[arXiv:nucl-th/9806004].
%%CITATION = NUCL-TH 9806004;%%
%%Cited 40 times in SPIRES-HEP

%\cite{Magas:1999yb}
\bibitem{Magas:1999yb}
V.~K.~Magas {\it et al.},
%``Kinetic freeze-out models,''
Heavy Ion Phys.\  {\bf 9}, 193 (1999)
[arXiv:nucl-th/9903045].
%%CITATION = NUCL-TH 9903045;%%
%%Cited 35 times in SPIRES-HEP

\bibitem{miklos05} T. Hirano and M. Gyulassy, nucl-th/0506049.

%\cite{Bass:1999tu}
\bibitem{Bass:1999tu}
  S.~A.~Bass, A.~Dumitru, M.~Bleicher, L.~Bravina, E.~Zabrodin, H.~St\"ocker and W.~Greiner,
  %``Hadronic freeze-out following a first order hadronization phase  transition
  %in ultrarelativistic heavy-ion collisions,''
  Phys.\ Rev.\ C {\bf 60}, 021902 (1999)
  [arXiv:nucl-th/9902062].
  %%CITATION = NUCL-TH 9902062;%%
%\cite{Dumitru:1999sf,Bass:2000ib}
\bibitem{Dumitru:1999sf}
A.~Dumitru, S.~A.~Bass, M.~Bleicher, H.~St\"ocker and W.~Greiner,
%``Direct emission of multiple strange baryons in ultrarelativistic  heavy-ion
%collisions from the phase boundary,''
Phys.\ Lett.\ B {\bf 460}, 411 (1999)
[arXiv:nucl-th/9901046].
%%CITATION = NUCL-TH 9901046;%%
%%Cited 17 times in SPIRES-HEP
%\cite{Bass:2000ib}
\bibitem{Bass:2000ib}
S.~A.~Bass and A.~Dumitru,
%``Dynamics of hot bulk QCD matter: From the quark-gluon plasma to  hadronic
%freeze-out,''
Phys.\ Rev.\ C {\bf 61}, 064909 (2000)
[arXiv:nucl-th/0001033].
%%CITATION = NUCL-TH 0001033;%%
%%Cited 73 times in SPIRES-HEP

\bibitem{teaney01} D. Teaney, J. Lauret, and E. Shuryak,
nucl-th/0110037.
%\cite{Nonaka:2005aj}
\bibitem{Nonaka:2005aj}
  C.~Nonaka and S.~A.~Bass,
  %``3-D hydro + cascade model at RHIC,''
  arXiv:nucl-th/0510038.
  %%CITATION = NUCL-TH 0510038;%%
  %%Cited 4 times in SPIRES-HEP
\bibitem{hirano05} T. Hirano, Proceedings of the Quark Matter 2005, Aug. 4 - 9, 2005,
Budapest, arXiv:nucl-th/0510005.

\bibitem{pasi01} P. Huovinen, private communications, (2003).

\bibitem{markusqm05} M. Oldenburg, {\it et al.} (STAR Collaboration),
Proceedings of Quark Matter 2005, Aug. 4 - 9, 2005, Budapest, arXiv:nucl-ex/0510026.

\bibitem{star1234} J. Adam {\it et al.} (STAR Collaboration), Phys.\ Lett.\ B {\bf 612}, 181 (2005).

\bibitem{Biro:1984}
T.S. Biro, H.B. Nielsen, J. Knoll,
Nucl.\ Phys.\ B {\bf 245}, 449 (1984).

%\cite{Mohring:1992wm}
\bibitem{Mohring:1992wm}
  H.~J.~Mohring, J.~Ranft, C.~Merino and C.~Pajares,
  %``String fusion in the dual parton model and the production of anti-hyperons
  %in heavy ion collisions,''
  Phys.\ Rev.\ D {\bf 47}, 4142 (1993).
  %%CITATION = PHRVA,D47,4142;%%

%\cite{Sorge:1995dp}
\bibitem{Sorge:1995dp}
  H.~Sorge,
  %``Flavor Production in Pb(160AGeV) on Pb Collisions: Effect of Color Ropes
  %and Hadronic Rescattering,''
  Phys.\ Rev.\ C {\bf 52}, 3291 (1995)
  [arXiv:nucl-th/9509007],
  %%CITATION = NUCL-TH 9509007;%%
  H.~Sorge, M.~Berenguer, H.~St\"ocker and W.~Greiner, Proceedings of Hot and Dense Nuclear Matter, 621-633, 1993, Bodrum.

\bibitem{erice03} N. Xu, Prog. Part. Nucl. Phys. {\bf 53}, 165(2004).




\end{thebibliography}
\end{document}